\documentclass[conference]{IEEEtran}
\IEEEoverridecommandlockouts
\usepackage{cite}
\usepackage{amsmath,amssymb,amsfonts}
\usepackage{algorithmic}
\usepackage{graphicx}
\usepackage{textcomp}
\usepackage{xcolor}
\usepackage{graphicx}
\usepackage{caption}
\usepackage{subcaption}
\usepackage{tikz}
\usepackage{graphicx}
\usepackage{caption}
\usepackage{subcaption}
\usepackage{booktabs}
\usepackage{booktabs,siunitx}
\usepackage[caption=false,font=footnotesize]{subfig}
\usepackage{booktabs,tabularx}
\usepackage[caption=false,font=footnotesize]{subfig}
\def\BibTeX{{\rm B\kern-.05em{\sc i\kern-.025em b}\kern-.08em

    T\kern-.1667em\lower.7ex\hbox{E}\kern-.125emX}}
\begin{document}

\title{Unsupervised Denoising of Real Clinical Low Dose Liver CT with Perceptual Attention Networks\\
\thanks{Shenzhen Science and Technology Innovation Program, Grant Number:
JCYJ20240813143102004

* He is the supervisor of this paper.

\# He is the corresponding author.}
}

\author{
\IEEEauthorblockN{1\textsuperscript{st} Zhilin Guan}
\IEEEauthorblockA{\textit{Department of Computing} \\
\textit{Harbin Institute of Technology}\\
Harbin, China \\
cysong@stu.hit.edu.cn}  

\and
\IEEEauthorblockN{2\textsuperscript{nd} Wei Zhang$^{*,\#}$}
\IEEEauthorblockA{\textit{Department of Computing} \\
\textit{Harbin Institute of Technology}\\
Harbin, China \\
wei.zhang.2@ki.se}
}
\maketitle


\begin{abstract}
With the development of deep learning, the medical image field has also been widely used it to assist in research, and the main research in this paper is to solve the low-dose computed tomography (LDCT) denoising problem using deep learning. Although LDCT reduces the radiation hazard to patients, it also brings more noise which has some visual interference to the doctor’s judgment and affects the diagnosis result. To solve this problem, referring to the architecture of cycle-generative adversarial networks (Cycle-GAN) for unsupervised learning, this paper innovatively proposes an end-to-end unsupervised LDCT denoising framework. It combines the U-Net structure for multi-scale feature extraction, the attention mechanism for feature fusion, the combination of residual network for feature transformation, and also consider the comparison of GAN network models and introduce perceptual loss to improve the network for the characteristics of medical images. In addition, we build a real LDCT database and design a large number of comparative experiments to validate the method, using both experimental standards in the image field and evaluation standards in the medical field. The main feature of this paper compared with classical methods is that this paper solves the drawback that real data cannot be used for supervised learning, while this experimental result still have quite excellent performance compared with classical excellent methods, which are professionally judged by imaging physicians and meet the clinical needs of physicians.
\end{abstract}


\section{Introduction}
CT technology is becoming an important tool for clinical diagnosis, but the radiation caused by X-rays in CT testing is a serious health hazard to the patient. In particular, women and children are more sensitive to radiation. The fragility of children’s organs makes them far more vulnerable than adults when undergoing CT scans and highly susceptible to cancer [1]. In the United States, radiation from CT examinations accounts for 67\% of the total radiation from medical examinations and continues to increase. Medical CT images are now ranked as the most important source of artificial radiation [2].

\begin{figure}[t]
  \centering
  \begin{subfigure}[b]{0.32\linewidth}
    \centering
    \includegraphics[width=\linewidth]{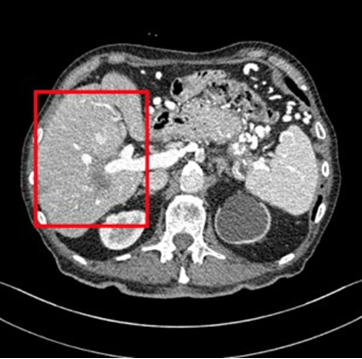}
    \caption{NDCT}
    \label{fig11}
  \end{subfigure}\hfill
  \begin{subfigure}[b]{0.32\linewidth}
    \centering
    \includegraphics[width=\linewidth]{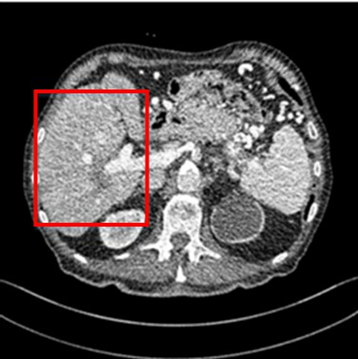}
    \caption{LDCT}
    \label{fig12}
  \end{subfigure}\hfill
  \begin{subfigure}[b]{0.32\linewidth}
    \centering
    \includegraphics[width=\linewidth]{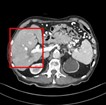}
    \caption{LDCT denoising}
    \label{fig13}
  \end{subfigure}
  \caption{The effect of the method proposed in this paper. (a) NDCT, the lesion in the region of interest is obvious; (b) LDCT, the lesion in the ROI is difficult to distinguish from the noise; (C) LDCT denosing with the method proposed in this paper, the lesion in the ROI is also clearly visible.}
  \label{fig1}
\end{figure}

Since radiation from CT tests is a serious health hazard to the patient, the ALARA (as low as reasonably achievable) guideline should be followed when performing CT tests on patients, i.e., the radiation dose during CT scans should be reduced as much as possible without affecting the clinical diagnosis. In recent years, LDCT has become a major research hotspot in the field of medical imaging. LDCT reduces the damage to the subject’s body by reducing the radiation intensity of X- rays [3, 4, 5], and reducing the dose of X-rays causes a large amount of noise and artifacts in CT, resulting in a reduction of CT image quality. These noise and artifacts often obscure important details in CT images, which may be potential lesion information, and the loss of these details will affect the clinical diagnosis and cause irreversible results [6]. Due to the negative impact of LDCT quality degradation on clinical diagnosis, the study of LDCT image denoising has received extensive attention in academic circles.

LDCT is not just about reducing the radiation dose as much as possible. With the reduction in radiation dose comes the possibility that the CT images may not support the physician’s diagnosis of the condition. The use of LDCT for diagnosis increases the difficulty of diagnosis, and may even cause doctors to make wrong judgments due to doctor's negligence or poor visualization of LDCT, or not finding the lesion, delaying the diagnosis of the patient and causing a series of serious consequences. It becomes extra important to process the LDCT visualization well [7, 8]. Some previous traditional methods are either poor or too slow, which will affect the doctor’s work efficiency and judgment of the disease to different degrees, so this paper intends to use deep learning technology to process the LDCT to enhance it and try to achieve the effect of NDCT to assist the doctor in the diagnosis of the disease as well as the review [9, 10]. Therefore the criteria for reducing the dose still have to meet the doctor’s diagnosis, because some organs are easier to observe, so the very mature LDCT technique can already be used for diagnosis, while there are more abdominal organs and smaller differences, and the abdomen is currently not available for low-dose technique, which is the main research of this paper. 

In order to obtain CT datasets with different doses requires different doses of CT detection on the subject, which will generate excessive radiation to the patient. In addition, patients inevitably breathe or jitter while undergoing CT examinations, making it difficult to obtain high-quality supervised datasets. For non-aligned CT datasets, supervised methods cannot be trained effectively, and current unsupervised CT denoising methods cannot effectively remove noise. To address the above problems, this paper proposes an end-to-end unsupervised LDCT denoising framework, which can ensure good denoising effect while preserving the detailed information in CT images more clearly. To achieve that goal, we are making a series of contributions covering all aspects of the problem. These contributions include:

(1)	We propose an end-to-end unsupervised LDCT denoising method, which combines a U-Net structure for multi-scale feature extraction, an attention mechanism for feature fusion, and a residual network for feature transformation.

(2)	We use the migration learning method for the application of 3D features to CT images. After training the 2D network, the parameters are then migrated so that the 3D information can be used to improve the experimental results.

(3)	To further improve the visual effect of LDCT, we add the perceptual loss to the loss function. The specific implementation is performed by using the VGG-19 for the extraction of perceptual features, taking the output of its lower 16 layers as the corresponding perceptual features.

\section{RELATED WORK}
Theoretically, the noise in LDCT mainly obeys the Poisson distribution, however, in the actual imaging results, there is usually a mixture of gaussian noise, pretzel noise and multiplicative noise. To address the problem of noise in LDCT images that adversely affects clinical diagnosis, researchers have proposed various methods for denoising LDCT images, and the proposed algorithms are divided into three main categories: projection domain pre-processing algorithms, iterative reconstruction algorithm, and image post-processing algorithms. Since the use of deep learning currently has an advantageous performance, and the methods using deep learning processing are post-processing methods, post-processing methods are currently more popular.

\subsection{Projection domain pre-processing algorithms}
The projection domain preprocessing algorithm is used to obtain the denoised CT image by denoising the projection domain data and then passing it through the CT reconstruction algorithm. These algorithms require the original projection domain data, which makes these methods more difficult to study. In addition, these algorithms often result in reduced resolution of CT images and blurred edges. The classical algorithms of projection domain preprocessing include structural adaptive filtering [11], bilateral filtering [12], and penalized weighted least squares [13], which often lead to the reduction of resolution and blurring of CT image edges. In recent years, sparse projection denoising [14] and penalized likeli-hood method [15] have been proposed by research scholars to reduce the blurring of CT image edges under the premise of CT image denoising in the projection domain.

\subsection{Iterative reconstruction algorithm}
Iterative reconstruction algorithms combine the data in the sinogram domain, prior information in the image domain, and even the statistical properties of the parameters of the imaging system into a unified objective function. Algorithms for iterative reconstruction [15, 16, 11] have all achieved good results, with different algorithms introducing different prior knowledge, using techniques such as full variational , dictionary learning, compressed sensing and its variants, non-local mean and physical acquisition. However, iterative reconstruction techniques with relatively high computational overhead, slow reconstruction speed and variation in image appearance limit their clinical application.

\subsection{Image post-processing algorithms}
Image post-processing algorithms directly denoise the reconstructed CT images, which can effectively suppress the noise and artifacts in LDCT images. Due to the good general applicability of image post-processing algorithms, it has become a hot research topic in the field of LDCT image denoising in recent years. The image post-processing methods are mainly include: adaptive nonlocal means filter, bilateral filtering, k-SVD denoising [17, 18], block-matching 3D [19-24] and methods based on deep learning[25-27]. Methods based on deep learning is a rapidly growing field with many practical applications.

\section{METHOD}
\begin{figure*}[t]
  \centering
  \includegraphics[width=\textwidth]{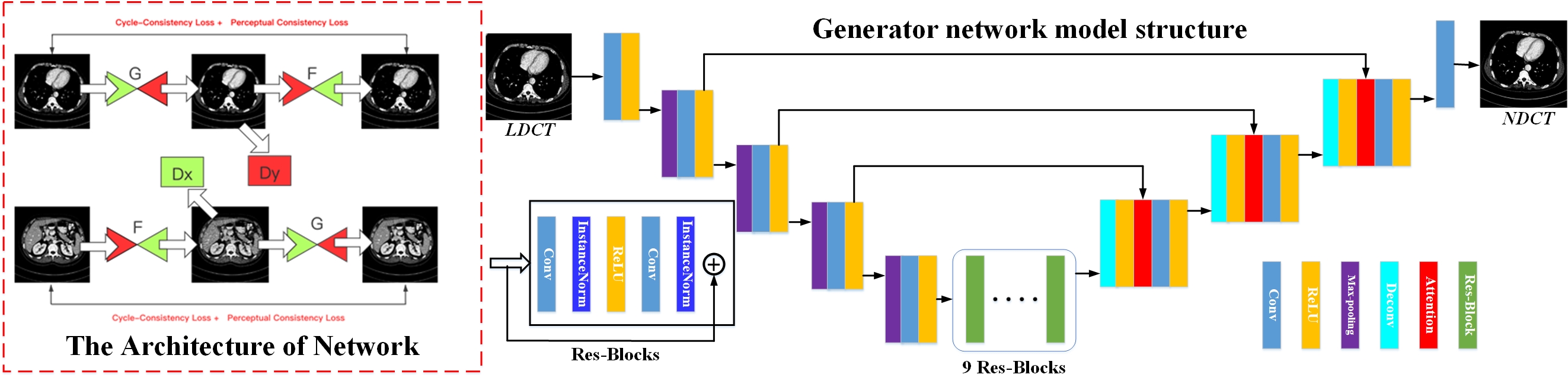}
  \caption{Proposed unsupervised LDCT denoising framework and generator architecture. The generator combines a U\mbox{-}Net structure for multi-scale feature extraction, an attention mechanism for feature fusion, and residual blocks for feature transformation.}
  \label{fig2}
\end{figure*}
In order to solve the LDCT denoising problem and apply our algorithm to real clinical data, we propose an unsupervised LDCT denoising algorithm based on deep learning. The major difference with the classical algorithm is that our algorithm is an unsupervised learning algorithm, and this paper refers to the idea of unsupervised image conversion with Cycle-GAN as the main body.
\subsection{Denoising model design}

Let $z \in \mathbb{R}^{N \times N}$ denote the low-dose CT (LDCT) image and  $x \in \mathbb{R}^{N \times N}$ the corresponding normal-dose CT (NDCT) image. The goal of the denoising process is to find a function $G$ that maps low-dose CT $z$ to standard metrology CT $x$:
\begin{equation}
D:\; z \to x
\end{equation}

On the other hand, we assume $z$ as the sample from the LDCT image distribution $P_{l}$ and $x$ from the normal metric distribution or the actual distribution$P_{r}$. The denoising function $G$ maps the samples from $P_{l}$ to a specific distribution $P_{g}$. By changing the function $G$, we change $P_{g}$ so that it is close to $P_{r}$. In this way, the denoising operator is moving one data distribution to another data distribution.
There is no clear clue how the data distributions of NDCT and LDCT are related to each other, which makes it difficult to denoise LDCT images using traditional methods. However, this uncertainty of noise can be ignored in deep learning denoising because deep neural networks extract high-level features from pixel data to represent the data distribution.

\subsection{The architecture of our network}	
Let $z \in \mathbb{R}^{N \times N}$ denote the low-dose CT (LDCT) image and  $x \in \mathbb{R}^{N \times N}$ the corresponding normal-dose CT (NDCT) image. The goal of denoising is to learn a mapping $G:\mathbb{R}^{N \times N}\to\mathbb{R}^{N \times N}$  such that $x \approx G(z)$.

The whole network is an end-to-end unsupervised LDCT denoising network, and the specific network structure is shown in the left half of Figure \ref{fig2}. In the Figure \ref{fig2}, $G$ denotes the generator of LDCT to NDCT, $F$ denotes the generator of NDCT to LDCT, DX and DY are the discriminators to determine whether it is NDCT and whether it is LDCT, respectively, while perceptual loss is also added in addition to the cyclic consistency loss. The network inputs a pair of CT images with different styles or contents as well, while initializing two discriminators $D_{X}$(determine whether it is a LDCT image), $D_{Y}$ (determine whether it is a NDCT image) and two generators $F$ and $G$, and training these four networks in a sequential loop in the generator-discriminator order.

Because U-Net has good effect in medical image field, we designed to use U-Net 256 in the generator that is downsampled 8 times, which can change a size of $256\times256$ to $1\times1$. Through experimental verification, it was indeed confirmed that U-Net works better. The analysis is mainly because U-Net combines shallow and deep features effectively and also allows multi-scale information collection, which retains more valuable information.

So, in the generator, the Encoder-Decoder structure similar to U-Net is added before and after the Resnet network. At the same time, a skip-connection structure is added for multi-scale fusion of features, and the attention module is used to fu- sion LDCT information and NDCT information with weights at the place of feature fusion, and nine Resnet structures are used for image conversion, so that feature information extraction, feature fusion, and image conversion are divided into three structures for operation, which is better than the U- net one structure is more satisfactory, and the later experiments also prove this conclusion. The structure of the generator is shown in the Figure \ref{fig2}: the LDCT image goes through Encoder for feature extraction and LDCT features are extracted, while the shallow LDCT features are passed to the later upsampling structure for feature fusion using the skip-connection structure, Decoder is used to recover from NDCT features to NDCT images, the 9 Res-Block structures in the middle are used for LDCT to NDCT feature conversion, and the attention structure achieves weighted fusion of LDCT information and NDCT information.
\begin{figure}[t]
  \centering
  \includegraphics[width=0.5\textwidth]{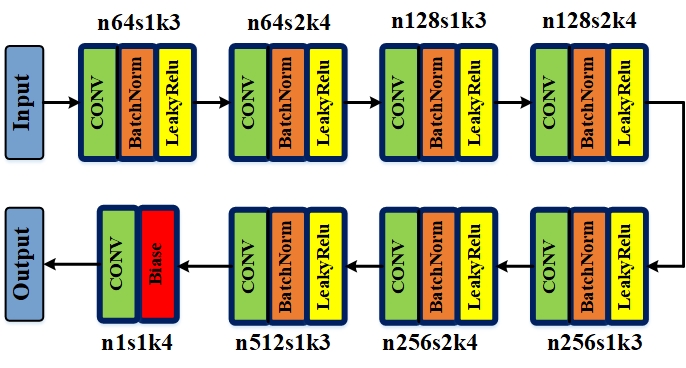}
  \caption{The discriminator uses the $70\times70$ discriminator struc- ture of PATCH-GAN for precise minutiae discrimination.}
  \label{fig3}
\end{figure}

In Figure \ref{fig3}, the discriminator uses the $70\times70$ discriminator structure of PATCH-GAN for accurate fine discrimination. The discriminator uses the $70\times70$ discriminator structure of PATCH-GAN for accurate fine discrimination. The discriminator has 8 convolutional layers, the first two layers have 64 convolutional kernels, then the next two layers will have 128 convolutional kernels, the next two layers will contain 256 convolutional kernels each, the last two layers have 512 convolutional kernels and 1 convolutional kernel and Biase layer respectively, and the last convolutional layer will have only a single output. A batch normalization layer is also added to the convolutional layers to stabilize/optimize the training of the GAN.

\subsection{Design of the Attention module}
\begin{figure}[h]
  \centering
  \includegraphics[width=0.5\textwidth]{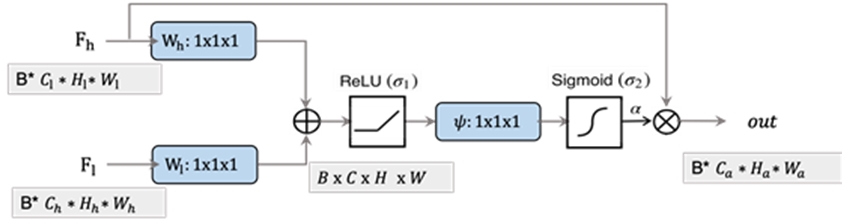}
  \caption{The attention structure.}
  \label{fig4}
\end{figure}

The purpose of the attention module in this paper is to make the LDCT information and NDCT information more effectively integrated, and the attention module is equivalent to giving a weight to LDCT features and NDCT features respecively. In the generator network, we use the U-Net structure, which is more suitable for the multi-scale feature extraction of medical images. In addition to the residual structure, we also add the Skip layer for the fusion of different features, and use the attention structure to eliminate some redundant information in the fusion process. The Attention structure is shown in Figure \ref{fig4}.
\begin{equation}
q(l,h) = \psi^{T}\,\sigma_{1}\!\left(W_h^{T} h_i + W_l^{T} l_i + b_l\right) + b_{\psi}
\end{equation}
\begin{equation}
a_i = \sigma_{2}\!\big(q(l,h)\big)
\end{equation}

$F_{l}$ is the feature of LDCT on the right side of the generator network, $F_{h}$ is the feature of NDCT on the left side of the  generator network, $W_{l}$ and Wh are the new feature information obtained by multiplying the calculated $a$ and $F_{h}$ by the convolutional layer with a convolutional kernel of 1, and $\sigma^2$ represents the Sigmoid function. Using this attention structure can achieve the effective fusion of feature information.

\subsection{Transfer learning using 3D information}
\begin{figure}[t]
  \centering
  \includegraphics[width=0.5\textwidth]{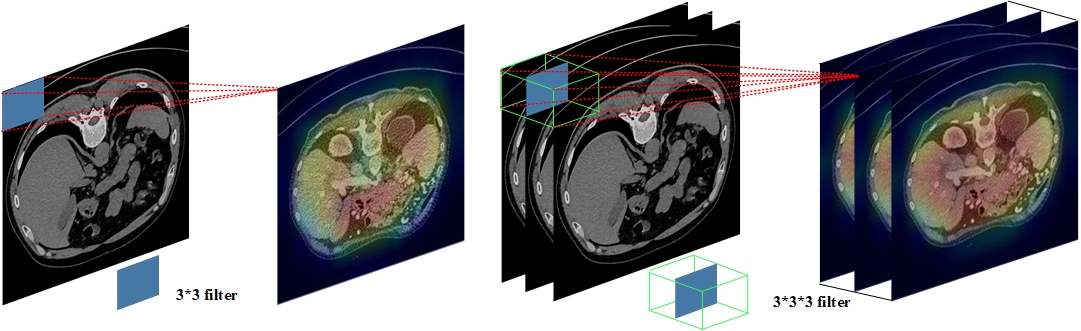}
  \caption{3D migration learning parameter sharing during LDCT denoising enhancement.}
  \label{fig5}
\end{figure}

We use transfer learning to accomplish parameters sharing and task migration. As shown in Figure \ref{fig5}, in terms of task migration, after we train the model of 80kv 1mm LDCT to 120kv 1mm NDCT conversion, we can train the 80kv 5mm LDCT to 120kv 5mm NDCT conversion model based on this model, which can reduce almost 80\% of the time than training a model from scratch, and similarly we can train the 100 kV 1 mm LDCT to 120kv 1mm NDCT conversion model. We adopt a 2.5D design that consumes three adjacent slices but performs slice wise inference. This choice balances contextual information and computational cost, keeps memory usage within the capacity of typical clinical GPUs, and preserves in plane resolution and latency stability. We deliberately avoid explicit through plane smoothing or full 3D convolutions in order to prevent unwanted blurring of thin vessels and to maintain a practical inference footprint. The model therefore does not enforce anatomical continuity along the z direction by design.

In terms of parameter sharing, after training the 2D model, we can migrate it to the 3D model by keeping the parameters, initializing the middle layer of the 3D model parameters to the 2D model parameters, and then setting the other parameters to 0, and then using migration learning for training. To utilize the 3D information, we feed three consecutive CTs at a time, the middle one is the LDCT to be transformed and the other two being its upper and lower adjacent LDCTs. In terms of network design, the original convolutional kernel is increased by one dimension, while the convolutional kernel of the middle layer can be initialized using the convolutional kernel parameters of the trained 2D model, and the parameters of the other two layers are initialized to 0. We adopt a 2.5D design that consumes three adjacent slices but performs slice wise inference. This choice balances contextual information and computational cost, keeps memory usage within the capacity of typical clinical GPUs, and preserves in plane resolution and latency stability. We deliberately avoid explicit through plane smoothing or full 3D convolutions in order to prevent unwanted blurring of thin vessels and to maintain a practical inference footprint. The model therefore does not enforce anatomical continuity along the z direction by design.

\subsection{Loss function}
$\mathcal{L}_{\mathrm{total}}$ consists of an adversarial term $\mathcal{L}_{\text{adv}}^{\text{gen}}$, a cycle-consistency term $\mathcal{L}_{\text{cyc}}$, and a perceptual-consistency term $\mathcal{L}_{\text{perc}}$.
The adversarial term drives the generated samples in both domains to match the corresponding real data distributions. The cycle-consistency term stabilizes content and structure in the unpaired setting. The perceptual term measures similarity in a deep feature space to enhance texture fidelity.

\begin{equation}
\mathcal{L}_{\text{total}}
= \lambda_{\text{adv}}\,\mathcal{L}_{\text{adv}}^{\text{gen}}
+ \lambda_{\text{cyc}}\,\mathcal{L}_{\text{cyc}}
+ \lambda_{\text{perc}}\,\mathcal{L}_{\text{perc}} .
\end{equation}

\begin{equation}
\mathcal{L}_{\text{adv}}^{\text{gen}}
= \mathbb{E}_{x\sim p_X}\,\ell\!\big(D_Y(G(x))\big)
+ \mathbb{E}_{y\sim p_Y}\,\ell\!\big(D_X(F(y))\big),
\end{equation}

\begin{equation}
\mathcal{L}_{\text{cyc}}
= \mathbb{E}_{x\sim p_X}\!\left[\lVert F(G(x))-x\rVert_{1}\right]
+ \mathbb{E}_{y\sim p_Y}\!\left[\lVert G(F(y))-y\rVert_{1}\right].
\end{equation}

\begin{equation}
\begin{split}
\mathcal{L}_{\text{perc}}
&= \sum_{\ell\in\mathcal{S}}\omega_{\ell}\Bigl(
\,\mathbb{E}_{x\sim p_X}\!\big[\|\phi_{\ell}(G(x))-\phi_{\ell}(y^\star(x))\|_2^2\big] \\
&\qquad\qquad\quad
+\,\mathbb{E}_{y\sim p_Y}\!\big[\|\phi_{\ell}(F(y))-\phi_{\ell}(x^\star(y))\|_2^2\big]
\Bigr).
\end{split}
\end{equation}

\section{EXPERIMENTAL RESULTS AND ANALYSIS}
This section contains the entire experimental part, from the collection and pre-processing of medical images, and then using different algorithms for experimental comparison, verifying the advantages of the algorithm in this paper. This algorithm is tested on two data sets to verify the effectiveness of the algorithm. At the same time, in addition to setting up medical evaluation indicators in this paper, it once again proves the effectiveness of the algorithm, which fully combines both medical and image aspects.

\subsection{Datasets for Training and Testing}
One of the data sets in this article is AAPM-Mayo. This data is the data set of the 2016 Low-Dose CT Challenge. The file types are DICOM and IMA formats. The standard measurement image data of this data set comes from real standard measurement data, and the low-dose image data is artificially added noise based on the standard measurement data by experts from the American Medical Association. This data has a one-to-one correspondence. The current LDCT denoising research uses this data set as the standard, so the supervised learning method is used to train the model. The other data set comes from Mudanjiang Second Affiliated Hospital. This data is real liver CT data, and the data format is DICOM format. The research content of this article is also mainly aimed at the research of real scene data. This data is the standard measurement data collected by different people and different periods. The characteristic of this data is that it does not have a one-to-one correspondence. At the same time, this data includes 1mm 80kv low-dose CT, 5mm 80kv low-dose CT, 1mm 100kv low-dose CT, 5mm 100kv low-dose CT, 1mm 120kv standard measurement CT, 3mm 120kv standard measurement CT.

First, the Catphan 500Plus body membrane is selected, and a cylinder with a low-contrast module diameter of 15 mm and a contrast ratio of 1.0\% was selected as the research object to simulate the low-contrast characteristics of the liver. Under the same scanning parameters, the tube voltage is 
120KV, tube current is 75mA, 100-400mA, automatic tube current adjustment (CARE Dose4D); tube voltage is 100KV, tube current is 75mA, 100-400mA, CARE Dose4D; tube voltage is 80KV, Tube current 75mA, 100-400mA, CARE Dose4D, scan the body membrane. Record the computed tomography dose index volume(CTDIvol) of each group of scanning plans, calculate the effective dose(ED), dose length product(DLP), and measure the image noise value(SD) and signal noise of the region of interest in each group of images Ratio(SNR), and two senior radiologists use a blind method to score the image quality of the region of interest. Compare the effects of different tube voltage and tube current conditions on CTD, ED, SD, SNR and image quality. The scan parameters of the low-dose experimental group were screened out through body membrane experiments.
\subsection{Data processing using window technology}
We use pydicom to identify and extract DICOM files and extract useful information inside DICOM files. After obtaining the image information of the DICOM file, window technology is used to add a preset window width and window level to the image data. The window level used in this article is 40 and the window width is 300.

The upper abdominal CT used in this paper uses a window level of 40 and a window width of 300, then the observable range is an interval with a width of 300 centered on 40, which is -110 to +190. The actual effects of window width and window level can be seen from Figure 6(a) and 6(b). The left figure does not increase the window width and window level, which is more difficult to observe. The right figure adds appropriate window width and window level, which is obviously more convenient for observation. At present, the pixel value range of the monitor we usually  use is 0-255, but generally the CT value will exceed this range, so in order to achieve visualization, it needs to be converted. The display range of the monitor is from  $y_{min}$ to  $y_{max}$,   $x$is the value of the input pixel, and  $y$ is the result after the change. 

\begin{figure}[h]
  \centering
  \begin{subfigure}[b]{0.5\linewidth}
    \centering
    \includegraphics[width=\linewidth]{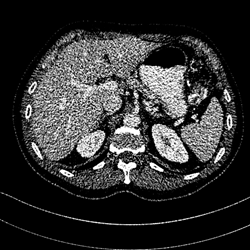}
    \caption{No window width and  level}
    \label{win1}
  \end{subfigure}\hfill
  \begin{subfigure}[b]{0.5\linewidth}
    \centering
    \includegraphics[width=\linewidth]{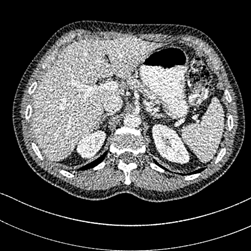}
    \caption{Set the window width and  level}
    \label{win2}
  \end{subfigure}\hfill
    \label{fig13}
  \caption{Visual comparison chart of window width and window position with and without window width.}
  \label{fig1}
\end{figure}

\begin{equation}
y =
\begin{cases}
y_{\min}, &
  x \le c - 0.5 - \dfrac{w-1}{2},\\[4pt]
\Bigl(\dfrac{x - (c - 0.5)}{\,w + 1\,} + 0.5\Bigr)\,(y_{\max}-y_{\min}), &
  \text{$x$ is other values}\\
y_{\max}, &
  x > c - 0.5 + \dfrac{w-1}{2}.
\end{cases}
\end{equation}

\subsection{Evaluation metrics}
The evaluation metrics used in this paper are not only tradi- tional image evaluation metrics, but also professional medi- cal evaluation metrics. The evaluation metrics for the AAPM dataset are PSNR, SSIM and perceptual loss . The evaluation metrics for the real dataset are the signal-to-noise ratio(SNR), the contrast noise ratio(CNR)and subjective evaluation(SE) metrics. Two experienced radiologist performed subjective evaluations of the collected CT images, which were evaluated on a 5-point scale. All CT images are evaluated using the same window width of 300 and window position of 40.
\subsection{Comparisons with State-of-the-Art Methods}
In the experimental results section, the experimental results of the methods used in this article will be shown in turn, as well as the comparison with other methods, and some non-training data will be added for testing to test the scalability of the algorithm.
\subsubsection{Classic method results in Mayo dataset}
First, it compares with three classic methods, one is WGAN-VGG [26], WGAN [27] and RED-CNN [23]. This experiment is conducted on the Mayo dataset. Through the comparison of the three methods, it can be found that Figure 7(d) RED-CNN has the problem of over-smoothing, while Figure 7(e) WGAN-VGG has a better visualization effect due to the addition of perceptual loss. It can also be found from the following indicators that the RED-CNN methods PSNR and SSIM have reached very high, but the perceptual loss is very large, and WGAN-VGG only has a better perceptual loss than the RED-CNN method, and it has a better good visualization effect. It can be seen that using perceptual loss in this problem will make the visualization effect better.

\begin{figure}[h]
  \centering
  \begin{subfigure}[b]{0.48\linewidth}
    \centering
    \includegraphics[width=\linewidth]{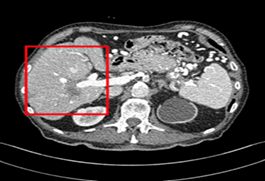}
    \caption{FULL-DOSE}
    \label{fig:full}
  \end{subfigure}\hfill
  \begin{subfigure}[b]{0.48\linewidth}
    \centering
    \includegraphics[width=\linewidth]{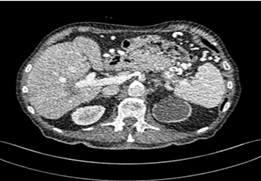}
    \caption{LOW-DOSE}
    \label{fig:low}
  \end{subfigure}


  \begin{subfigure}[b]{0.32\linewidth}
    \centering
    \includegraphics[width=\linewidth]{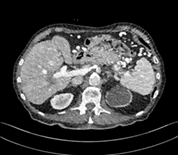}
    \caption{WGAN}
    \label{fig:wgan}
  \end{subfigure}\hfill
  \begin{subfigure}[b]{0.34\linewidth}
    \centering
    \includegraphics[width=\linewidth]{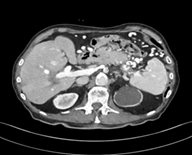}
    \caption{RED-CNN}
    \label{fig:redcnn}
  \end{subfigure}\hfill
  \begin{subfigure}[b]{0.32\linewidth}
    \centering
    \includegraphics[width=\linewidth]{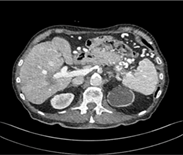}
    \caption{WGAN-VGG}
    \label{fig:wganvgg}
  \end{subfigure}

  \caption{Comparison of classic methods.}
  \label{fig:five-grid}
\end{figure}

\begin{table}[h]
  \centering
  \caption{Comparison of classic methods}
  \label{tab:classic-mayo}
  \begin{tabular}{lccc}
    \toprule
    \textbf{Methods} & \textbf{PSNR} & \textbf{SSIM} & \textbf{PL} \\
    \midrule
    LDCT      & 26.068 & 0.8440 & 4.78 \\
    RED-CNN   & \textbf{31.418} & \textbf{0.9701} & 4.26 \\
    WGAN      & 30.166 & 0.9027 & 4.02 \\
    WGAN-VGG  & 29.715 & 0.8984 & \textbf{2.33} \\
    \bottomrule
  \end{tabular}
\end{table}
\subsubsection{Our method results on the Mayo dataset.}
This paper proposes to use the CYCLE-GAN architecture as the framework to conduct experimental comparisons of different generator designs. The best solution is the generator network proposed in this paper. Figure 8 compares the experimental results of the methods proposed in this article, and proves the effectiveness of this method on the AAPM-Mayo data set. The method proposed in this article has better effects than ordinary CYCLE-GAN and is more suitable for this dataset.  At the same time, it can be seen in Table \ref{tab:ours-mayo-ieee} that the effect of the method in this paper is indistinguishable from the WGAN-VGG method, but for an unsupervised learning method, the effect is already very good.

\begin{figure}[h]
  \centering
  \begin{subfigure}[b]{0.48\linewidth}
    \centering
    \includegraphics[width=\linewidth]{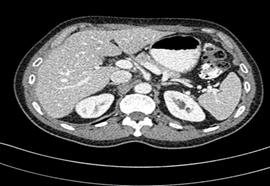}
    \caption{FULL-DOSE}
    \label{fig:full}
  \end{subfigure}\hfill
  \begin{subfigure}[b]{0.48\linewidth}
    \centering
    \includegraphics[width=\linewidth]{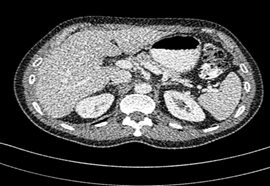}
    \caption{LOW-DOSE}
    \label{fig:low}
  \end{subfigure}
  
  
  \begin{subfigure}[b]{0.32\linewidth}
    \centering
    \includegraphics[width=\linewidth]{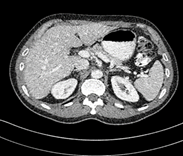}
    \caption{CYCLEGAN-RES	}
    \label{fig:wgan}
  \end{subfigure}\hfill
  \begin{subfigure}[b]{0.32\linewidth}
    \centering
    \includegraphics[width=\linewidth]{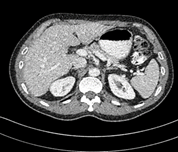}
    \caption{CYCLEGAN-U}
    \label{fig:redcnn}
  \end{subfigure}\hfill
  \begin{subfigure}[b]{0.32\linewidth}
    \centering
    \includegraphics[width=\linewidth]{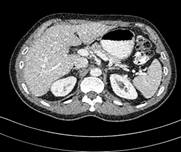}
    \caption{CYCLEGAN-RAS}
    \label{fig:wganvgg}
  \end{subfigure}
  \caption{Experimental results of the method in this paper}
  \label{fig:five-grid}
\end{figure}

\begin{table}[h]
  \caption{Our method results on Mayo dataset}
  \label{tab:ours-mayo-ieee}
  \centering
  \begin{tabular}{p{0.52\linewidth}ccc}
    \hline
    \textbf{Methods} & \textbf{PSNR} & \textbf{SSIM} & \textbf{PL} \\
    \hline
    LDCT & 26.068 & 0.844 & 4.78 \\
Perceptua+Res-Net & 28.765 & 0.8964 & 3.17 \\
Perceptual+U-Net   & 28.981 & 0.9001 & 2.94 \\
Perceptual+ResNet-Attention+skip
                                   & \textbf{29.176} & \textbf{0.9076} & \textbf{2.63} \\
    \hline
  \end{tabular}
\end{table}
\subsubsection{Our method results in real dataset}
From the Figures (d) and (e) in Figure 8 and the data in the Table \ref{tab:ours-mayo-ieee}, it can be seen that although the PSNR and SSIM indicators of RED-CNN are particularly high and are much higher than WGAN-VGG, the PL value of WGAN-VGG is much smaller than RED-CNN. RED-CNN also produces an over-smoothing effect, and the deviation from the real data is too large, which shows the importance of the PL index. The peak signal-to-noise ratio has always been an important criterion for evaluating picture quality. It measures the gap between pixels and is based entirely on the similarity evaluation of pixels. However, because it is only sensitive to pixels and does not conform to the observation rules of the human eye, the human eye does not observe in pixels. This evaluation lacks integrity and understanding of the content. Therefore, the peak signal-to-noise ratio cannot be used to judge the quality of the image. Generally speaking, if the peak signal-to-noise ratio is greater than 30, the observation effect is better. Compared with the classic method, the method in this paper has improved indicators.
\begin{minipage}[t]{0.48\textwidth}
\centering
\captionof{table}{CTDIvol(mGy) statistics for 20 patients}
\resizebox{\linewidth}{!}{%
\begin{tabular}{l*{10}{S[table-format=2.2]}}
\toprule
 & {1}&{2}&{3}&{4}&{5}&{6}&{7}&{8}&{9}&{10}\\
\midrule
120 kV & 20.32 & 17.73 & 17.30 & 15.16 & 15.65 & 13.80 & 11.61 & 13.97 & 17.01 & 16.99 \\
80  kV & 12.47 & 11.87 & 10.00 &  9.73 & 10.58 & 10.02 &  8.94 &  7.94 &  8.72 & 10.10 \\
\bottomrule
\label{Table3}
\end{tabular}}
\end{minipage}

\begin{minipage}[t]{0.48\textwidth}
\centering
\captionof{table}{CTDIvol(mGy) statistics for 20 patients}
\resizebox{\linewidth}{!}{%
\begin{tabular}{l*{10}{S[table-format=2.2]}}
\toprule
 & {1}&{2}&{3}&{4}&{5}&{6}&{7}&{8}&{9}&{10}\\
\midrule
 120 kV & 24.32 & 21.73 & 21.30 & 19.16 & 19.65 & 17.80 & 15.61 & 17.97 & 21.01 & 20.99 \\
80  kV & 16.47 & 15.87 & 14.00 & 13.73 & 14.58 & 14.02 & 12.94 & 11.94 & 12.72 & 14.10 \\
\bottomrule
\label{Table4}
\end{tabular}}
\end{minipage}

\begin{minipage}[t]{0.48\textwidth}
\centering
\captionof{table}{CTDIvol(mGy) statistics for 20 patients}
\resizebox{\linewidth}{!}{%
\begin{tabular}{l*{10}{S[table-format=2.2]}}
\toprule
 & {1}&{2}&{3}&{4}&{5}&{6}&{7}&{8}&{9}&{10}\\
\midrule
 120 kV & 0.36 & 0.33 & 0.32 & 0.29 & 0.29 & 0.27 & 0.23 & 0.27 & 0.32 & 0.31 \\
 80  kV & 0.25 & 0.24 & 0.21 & 0.21 & 0.22 & 0.21 & 0.19 & 0.18 & 0.19 & 0.21 \\
\bottomrule
\label{Table5}
\end{tabular}}
\end{minipage}

To analyze the radiation dose variation patterns under different scanning voltages, the CTDIvol, DLP, and ED values at 120 kV and 80 kV were statistically analyzed. The results are shown in Table \ref{Table3}–Table \ref{Table5}.
Table \ref{Table3} shows the statistical results for the Computed Tomography Dose Index (CTD). As can be seen from the table, the CTD values at both voltages exhibit a certain fluctuation trend with increasing scan number. However, overall, the CTD values at 120 kV are significantly higher than those at 80 kV, indicating that higher tube voltages lead to an increase in absorbed dose per unit volume.
Table \ref{Table4} shows the statistical results for the Dose Length Product (DLP). The results show that DLP values at 120 kV are generally higher than those at 80 kV, with the same trend as the CTD values. This indicates that, given the same scan length, higher voltages result in a higher accumulated radiation dose.
Table \ref{Table5} shows the statistical results for the Effective Dose (ED). It can be observed that ED values at 80 kV are generally lower than those at 120 kV, indicating that lowering the tube voltage can effectively reduce the patient's overall radiation dose. Taking all three indicators into consideration, 80 kV scanning significantly reduces radiation exposure while maintaining image quality, which has positive implications for radiation protection.

\subsubsection{Universality test results}
In addition to the conversion from the low-dose training 80kv to the standard 120kv, this algorithm also adds an experiment from the low-dose 100kv to the standard 120kv. Also conducted an 80kv 5mm low-dose CT denoising experiment. According to the doctor's visual judgment, the effect of denoising experiments is not obvious. The results are as follows:

\begin{figure}[t]
  \centering
  \begin{subfigure}[b]{0.32\linewidth}
    \centering
    \includegraphics[width=\linewidth]{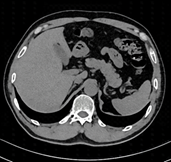}
    \caption{100 kV, 1 mm}
  \end{subfigure}\hfill
  \begin{subfigure}[b]{0.32\linewidth}
    \centering
    \includegraphics[width=\linewidth]{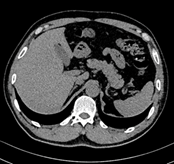}
    \caption{fake 120 kV, 1 mm}
  \end{subfigure}\hfill
  \begin{subfigure}[b]{0.31\linewidth}
    \centering
    \includegraphics[width=\linewidth]{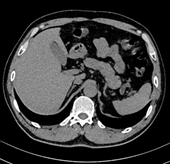}
    \caption{real 120 kV, 1 mm}
  \end{subfigure}
  \caption{Low-dose 100 kV 1mm experimental effect }
  \label{fig:ld_100kv_1mm}
\end{figure}

\begin{figure}[t]
  \centering
  \begin{subfigure}[b]{0.335\linewidth}
    \centering
    \includegraphics[width=\linewidth]{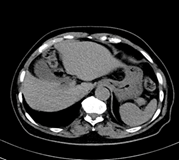}
    \caption{100 kV, 1 mm}
  \end{subfigure}\hfill
  \begin{subfigure}[b]{0.31\linewidth}
    \centering
    \includegraphics[width=\linewidth]{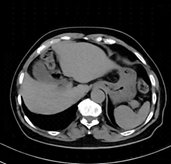}
    \caption{fake 120 kV, 1 mm}
  \end{subfigure}\hfill
  \begin{subfigure}[b]{0.32\linewidth}
    \centering
    \includegraphics[width=\linewidth]{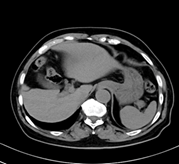}
    \caption{real 120 kV, 1 mm}
  \end{subfigure}
  \caption{Low-dose 80kV 5mm experimental results}
  \label{fig:ld_100kv_1mm}
\end{figure}

\section{CONCLUSION}
The problem addressed in this paper is the study of algorithms for the practical application of denoising of LDCT. Previous research works use supervised learning methods, but they do not work in clinical practical data. This paper innovatively proposes an end-to-end unsupervised LDCT denoising framework. It combines the U-Net structure for multi-scale feature extraction, the attention mechanism for feature fusion, the combination of residual network for feature transformation, and also introduce perceptual loss to improve the network. We use migration learning to accomplish parameter sharing and task migration, saving training time, making fuller use of the 3D spatial information of CT, and reducing the computational power requirements. This paper also contributes a dataset with NDCT and LDCT, it is real and valid, and this dataset can be used for further studies in the future. The main feature of this paper in comparison with classical methods is that we use unsupervised learning to address the limitation that real data cannot be used for supervised learning. The experimental results still yield excellent performance compared to classical methods, and our methods are excellent for real physician needs.

\section*{Acknowledgment}

This work is partly funded by Shenzhen Science and Technology Innovation Program via grant JCYJ20240813143102004.

\end{document}